 \documentclass[preprint,11pt]{elsarticle}

 \usepackage{graphics}

\usepackage{amssymb}
\usepackage{amsthm}
\usepackage{amsmath}

\journal{Some Journal}

\begin{document}
\newtheorem{remark}{\textbf{Remark}}
\newtheorem{theorem}{\textbf{Theorem}}
\newtheorem{definition}{\textbf{Definition}}
\newtheorem{proposition}{\textbf{Proposition}}
\newtheorem{corollary}{\textbf{Corollary}}

\begin{frontmatter}



 \title{Active margin system for margin loans and its application in
Chinese market: using cash and
randomly selected stock as collateral}



\author{Guanghui Huang\corref{cor1}}
\ead{hgh@cqu.edu.cn}

\author{ Wenting Xing }

\author{Weiqing Gu }

\cortext[cor1]{Corresponding author.}
\address{College of Mathematics and Statistics, Chongqing University,
  Chongqing 401331, China}

\begin{abstract}
An active margin system for margin loans is proposed for Chinese margin lending market, which uses cash and randomly selected stock as collateral. The conditional probability of negative return(CPNR) after a forced sale of securities from under-margined account in a falling market is used to measure the risk faced by the brokers, and the margin system is chosen under the constraint of the risk measure. In order to calculate CPNR, a recursive algorithm is proposed under a Markov chain model, which is constructed by sample learning method. The resulted margin system is an active system, which is able to adjust actively with respect to the changes of stock prices and the changes of different collateral. The resulted margin system is applied to 30,000 margin loans of 150 stocks listed on Shanghai Stock Exchange. The empirical results show the number of margin calls and the average costs of the loans under the proposed margin system are less than their counterparts under the system required by SSE and SZSE.
\end{abstract}


\end{frontmatter}



\section{Introduction}

 Margin loan is a loan using securities as collateral to get credit from a broker. A margin loan gives investors the possibility to buy more securities than normal and in that way increase profits or losses \cite{fortune2000,fortune2001,ricke2003}. Margin lending had been prohibited in China from 1996 to 2009, because of the absence of regulation on Shanghai Stock Exchange and Shenzhen Stock Exchange(SSE and SZSE). On March 31,2010, the margin lending markets have been established at SSE and SZSE respectively. With restarting the margin trading in Chinese market, how to set the scientific and effective regulatory system for the margin lending market becomes the question that the theorists and practitioners concern about. The development of margin lending markets in China motives this research for risk measurement and margin system construction for margin lending transactions.

 Although margin lending market provides the possibility to make the financial market more efficiently, it has been regarded as one of the source of instability of stock market for a long time \cite{huang2010}. The low margin requirement is one of the major contributing factors leading to the Stock Market Crash of 1929, which in turn contributed to the Great Depression, a troubling financial time in the 1930s\cite{galbraith}.

  In 1934, The United States Congress gave the Federal Reserve Board (FRB) the power to control initial margin requirement for the purposes of restraining the level of speculation and protecting investors and brokers \cite{fortune2000}. The initial margin requirement was changed 22 times between 1934 and 1974, and it has been fixed at 50\% since 1974 \cite{fortune2000}. The stock market crash of 1987 and the stock market boom in the late 1990s led to calls for the return to an active margin policy which sets a minimum equity position on the date of a credit-financed security transaction \cite{ricke2003,shiller2000}. The purpose of this paper is to set an active margin system for margin loans traded in Chinese market.

 Pursuant to the regulations of SSE and the SZSE, the initial and maintenance margin ratios are 50\% and 130\% respectively,which is called the required margin system in this paper. Recommendations for margin requirements are given by supervising authorities, it is not apparent how these recommendations are derived \cite{ argiriou2009}. However, a risk measurement based on conditional probability is proposed in this paper, which is used to set margin system with a specified target of risk control.

From the perspective of the  broker, the worst case is that the investors default once receive margin call, negative return still appears through mandatory liquidation. With regards to this, it is supposed that the investors will default whenever they receive margin call for the first time. The probability of negative return under the condition that the first margin call has been issued can describe the risk of a margin lending transaction, which is called the conditional probability of negative return(CPNR) in the following sections. CPNR does not only take into account the possibility of customer default, but also take into account the dynamic of the prices of collateral when mandatory liquidation is exercised  \cite{huang2010}. According to the regulations of two Chinese markets, the brokers must sell the collateral within two trading days, when the investors do not meet their margin call in time, and it does not know whether the margin trading can make a positive return until the mandatory liquidation is exercised.

The optimal margin system should be to minimize the chance cost, to minimize the rate of default, to minimize the frequency of margin calls \cite{huang2010}. With the help of CPNR, the resulted margin system can be explained more accurate than those margin systems just following the recommendations of the authorities. These factors including the amount of initial margin, the required maintenance margin ratio, the current stock price, the historical evolution of the stock prices, the riskless interest rate and the rate of margin loan can determine the value of CPNR.

 In order to calculate the value of CPNR, a mathematical model of stock price should be proposed. Markov chain model used as the working model to describe the characteristics of stock price process is appropriate \cite{huang2010}. A recursive algorithm is proposed to calculate CPNR in the following sections.

 A simple margin loan is considered in this paper, where an amount of cash and a portion of the randomly selected stock to be purchased are used as collateral. An active margin system is given for a particular margin account, such that the value of CPNR is less than or equal to the desired value of broker. The margin system deduced from the constraint of CPNR is called the deduced margin system in this paper.

 The resulted margin system is such a margin system chosen from the indifferent set with CPNR=0.05 using a least squares method and is applied to 153 stocks listed in Shanghai Stock Exchange, and for each stock there are 200 margin lending transactions with 30-day period. The frequency of negative return is observed for each stock under the deduced margin system, and it is found that there are 150 stocks whose frequency are less than 5\%, which are said to pass the out-of-sample test.

 A more detailed analysis is applied to those 150 stocks passing the out-of-sample test. The average initial margin ratio of 30,000=150$\times$200 margin loans is around 55\%, the average maintenance margin ratio is around 112\%, the average ratio of the deposited stock to the total mortgage is around 51\%. The number of margin call for each stock within 200 margin loans is around 19.49 under the required minimal margin system, and its counterpart under the deduced margin system  is averagely 5.29 for each stock. The cost of each margin loan is averagely 9.56 yuan under the required minimal margin system, and its counterpart under the deduced margin system is 9.15 yuan. In other words, the frequency of margin call is reduced by 73\% with less 4\% cost under the deduced margin system. The empirical results show that the active margin system deduced by CPNR is an operational margin system.

The outline of this paper is organized as follows. The risk measurement in Chinese margin lending market is discussed in Section 2. The algorithm to realize this risk measurement using a Markov chain as the working model is given in Section 3. An active margin system is constructed for the loans using cash and randomly closed stock as collateral in Section 4. Section 5 gives the empirical analysis of Chinese Market. Finally, the discussions and conclusions are given in Section 6.

\section{Risk measure for Chinese market}

\subsection{Margin system of Chinese market}

China has began to prepare the work of the margin trading since 2006. On January 8,2010, the state council agreed in principle to set up pilot margin trading, and recommended that the initial margin ratios of the two Chinese stock exchanges can not be less 50\%, the maintenance margin ratio of two stock exchanges can not be less 130\%, where the initial margin ratio,which is the ratio of initial margin to the market value of the stocks bought by the loan, the maintenance margin ratio,which is the ratio of the market value of the collateral to the market value of the stocks bought by the loan.

The maintenance margin ratio is marked-to-market by the broker, the investor under-margined account would face the phenomenon of a forced sale of stock in a falling market. That is the broker should issue a margin call when the current value of maintenance margin ratio is below the required minimal maintenance margin ratio, if the investors do not meet their margin call in time, the broker can sell collateral to meet the margin call, which is called mandatory liquidation in Chinese margin lending market.

\subsection{Conditional Probability of Negative Return }

In order to quantify the process of risk measurement for Chinese margin lending transaction, a simple margin loan using cash and randomly selected stock from 150 stocks investigated in this paper as collateral is constructed in the remainder of section. Denote the amount of initial margin as $Q_0+\delta P_0^{'}$ yuan, where $Q_0$ is the amount of cash, $P_0^{'}$ is the market price of the deposited stock on the date of the transaction, $\delta$ ($0\leq\delta\leq1$) is the proportion of the stock.The initial margin ratio on the date of margin lending is
\begin{equation}\label{initial_ratio}
m_{0}=\frac{Q_0+\delta P_0'}{P_0} ,
\end{equation}
where $P_i^{'}$ and $P_i$ is the market price of the deposited stock and the purchased stock for the $ith$ trading day respectively $(i=1,2,\ldots,T)$, where $T$ is the expiration date of the loan. And the maintenance margin ratio on the date of trading is

\begin{equation}\label{initial_ratio}
\omega_{0}=\frac{m_{0}P_{0}+ P_0}{P_0}=m_{0}+1.
\end{equation}

\textbf{Proposition 1 }.
Suppose the initial margin ratio on the date of the margin lending is $m_{o}$,the required minimal maintenance margin ratio is $\omega$, then
\begin{equation}
\frac{m_{0}P_0+P_0}{P_0}\geq\omega.
\end{equation}
Otherwise, the broker should ask the investor to deposit more margin to fulfill the requirement of maintenance margin.

The cash in initial margin can be divided into two parts on the $ith$ day, including the required amount of initial margin and the remaining margin after fulfilling the required initial margin,i.e.
\begin{equation}
Q_0(1+r)^i=\Sigma_i+ L_i,
\end{equation}
the required amount of initial margin is
\begin{equation}
 \Sigma_i=\omega P_0(1+R)^{i}-P_i-\delta P_i^{'},
\end{equation}
 the remaining margin after fulfilling the required initial margin is
\begin{equation}
 L_i=Q_0(1+r)^{i}-\omega P_0(1+R)^{i}+P_i+\delta P_i^{'},
\end{equation}
where $r$ is the risk free interest rate for one day, $R$ is the loan interest rate for one day. In the following sections, for simplicity, suppose that the interest rate of the loan is the same as the riskless interest rate.

The broker should issue a margin call, once there is no more remaining margin, i.e.
\begin{equation}
L_\tau=Q_0(1+r)^{\tau}-\omega P_0(1+r)^{\tau}+P_\tau+\delta P_\tau^{'}\leq0,
\end{equation}
where $\tau$ is the stopping time of the first margin call during the period of the loan, i.e.
\begin{equation}
\tau=min\{i\in\{1,2,\ldots,T\}:Q_0(1+r)^{i}-wP_0(1+r)^{i}+P_i+\delta P_i^{'}\leq0\}.
\end{equation}

The regulation of the two Chinese markets shows that the collateral should be liquidated within two trading days, once the investors do not meet the margin call in time. For simplicity, suppose that the investors would default whenever they receive a margin call, and the collateral should be liquidated within one trading day, so the day of mandatory liquidation $\tau^{^{*}}$ can be written as
\begin{equation}
\tau^{*}=min\{\tau+1,T\}.
\end{equation}

The return of the margin lending transactions after the collateral is liquidated is
\begin{equation}
Q_0(1+r)^{\tau^{*}}+P_{\tau^{*}}+\delta P_{\tau^{*}}^{'}-P_0(1+r)^{\tau^{*}}.
\end{equation}
The condition of a margin call and the negative return can be written as respectively
\begin{equation}
C=\{P_\tau+\delta P_\tau^{'}\leq\omega P_0(1+r)^{\tau}-Q_0(1+r)^{\tau}\},
\end{equation}

\begin{equation}
N=\{p_{\tau^{*}}+\delta P_{\tau^{*}}^{'}<P_0(1+r)^{\tau^{*}}-Q_0(1+r)^{\tau^{*}}\}.
\end{equation}

Where $C$ denote the event that the broker issues a margin call for the first time within the loan period, and $N$ denote the event that the margin lending transaction brings the broker a negative return after mandatory liquidation.

The conditional probability of negative return (CPNR) can be denoted as

\begin{equation}
CPNR=Prob\{N|C\}.
\end{equation}
where $Prob\{N|C\}$ is the conditional probability of $N$ given $C$. The value of CPNR is calculated under a Markov chain model in the following sections.

\section{CPNR under Markov Chain}
\subsection{Formation of the customer collateral combination}
Because it is not easy to get the real data of Combination of collateral, the data is got by random sampling method in this paper. According to the list of mortgage-backed securities and the list of underlying securities in Chinese two stock exchanges, the stock price data can be collected from the database of mortgage-backed securities and the underlying securities.

In the empirical study, it is necessary to judge whether the Combination of collateral has enough ability for margin, there are some securities chosen from the database of mortgage-backed securities to constitute the combination of collateral. From the point of view of the opportunity cost, it does not only know the types and the relative shares of the securities which is from the combination of collateral, but also know the best positions during carrying on margin lending transaction.

In this paper, a simple margin loan using cash and randomly selected stock to buy only one share of stock from 150 stocks investigated in this paper is constructed in the following sections.

\subsection{Construction of Markov chain}

The value of CPNR is calculated under a Markov chain model and the sequence of stock price is a Markov chain \cite{huang2010}. A Markov chain is a random process that jumps from one state to another. In order to construct a Markov model, the daily closing stock prices are used to estimate the state space and the transition probabilities. The combination of collateral is regarded as an index in this paper, the observed index prices are arranged in increasing order, every $g$ different prices are regarded as one state, obtaining the different states of the observed price data. Suppose there are $n$ elements in the state space, denoted as $S=\{s_1,s_2,\cdots,s_n\}$. Next, the work is to estimate transition probability between states, because the state space of the observed price data is fixed, the estimation of transition probabilities between states is more straightforward.

Denote the number of transition from state $i$ to state $j$ after one step as $f_{ij}$, the transition probability from the state $i$ to the state $j$ as $\hat{p}_{ij}(1)$, the total number of the index price falling into states $i$ as $f_{i.}$ , i.e.
\begin{equation}
\hat{p}_{ij}(1)=\frac{f_{ij}}{f_{i\cdot}},
\end{equation}
where $f_{i.}=\Sigma_{i=1}^{n}f_{ij}$ , and  $\hat{p}_{ij}(1)$ is the element of the one step transition matrix $P(1)$, $(i\in\{1,\cdots,n\},j\in\{1,\cdots,n\})$.

Suppose the chain is stationary, so the $n-step$ transition matrix $P(n)$ can be estimated by $P(n)=P^{n}(1)$.

\subsection{Stopping time and CPNR}

The conditional probability of negative return(CPNR) is rewritten as
\begin{equation}
Prob\{CPNR\}=prob\{N|C\}=\frac{Prob\{NC\}}{Prob\{C\}}.
\end{equation}

It shows that the $Prob\{C\}$ is non zero. On the other hand, when the $Prob\{C\}$ is zero, the value of CPNR is also zero by its definition. Denote the event that the broker issues a margin call on the $ith$ day as $D_i$, i.e.
\begin{equation}
  D_i=\{ P_i+\delta P_i^{'}\leq(w P_0-Q_0)(1+r)^i\},
\end{equation}
where $i=1,2,\cdots,T$.

In order to calculate CPNR, $Prob\{NC\}$ and $Prob\{C\}$ should be calculated by the following proposition.

\textbf{Proposition 2 (A recursive algorithm for \textbf{$prob\{C\} $})}.
Denote the event that the broker issues a margin call on the $ith$ day for the first time within the loan period as $C_t$,
suppose the current stock price $P_0+\delta P_0^{'}$ belong to the state space $s_h$, where $h\in\{1,2,\cdots,T\}$. Then $Prob(C)$ is given by
\begin{equation}
Prob(C)=\Sigma_{t=1}^{T}Prob\{C_t\},
\end{equation}
and
\begin{equation}
\mathrm{Prob}(C_t)=\mathrm{Prob}(\overline{D}_1)
\mathrm{Prob}(\overline{D}_2|\overline{D}_1)\cdots{\mathrm{Prob}(\overline{D}_{t-1}|
\overline{D}_{t-2})}\mathrm{Prob}({D_{t}}|\overline{D}_{t-1}),
\end{equation}
where

\begin{equation}
\left\{ \begin{aligned}
         \mathrm{Prob}(\overline{D}_1)&=1-\sum_{i=1}^{k_1}\hat{p}_{hi}(1),\\
         \mathrm{Prob}(D_m|\overline{D}_{m-1})&
         =\frac{\sum_{i=k_{m-1}+1}^n
                  \sum_{j=1}^{k_m}\hat{p}_{hi}(m-1)
                   \hat{p}_{ij}(1)}{\sum_{i=k_{m-1}+1}^{n}
                    \hat{p}_{hi}(m-1)},\\
  \mathrm{Prob}(\overline{D}_m|\overline{D}_{m-1}) &
         = 1-\mathrm{Prob}(D_m|\overline{D}_{m-1}),\\
         \end{aligned} \right.
                          \end{equation}

$m=2,\cdots,t$, and $k_m$ is the largest state index which satisfies
\begin{equation}
k_m=max\{k\in\{1,2,\cdots,n\}:S_{k}<(wp_0-Q_0)(1+r)^m\},
\end{equation}
where $S_{k}$ is the representative price level for the $kth$ state.

\textbf{Proposition 3 (A recursive algorithm for \textbf{$prob\{NC\}$} )}.  Suppose the current stock price $P_0+\delta P_0^{'}$ belong to the state space $s_h$, where $h\in\{1,2,\cdots,T\}$. then the probability that the two events $N$ and $C$ have happened at the same time is given by
\begin{equation}
Prob(NC)=\Sigma_{t=1}^T Prob(NC_t)£¬
\end{equation}
where
\begin{equation}
\begin{aligned}
         \mathrm{Prob}(NC_t)= &\mathrm{Prob}
         (\overline{D}_1)\mathrm{Prob}(\overline{D}_2|\overline{D}_1)
         \cdots \\
         &
         {\mathrm{Prob}(\overline{D}_{t-1}|
          \overline{D}_{t-2})}\mathrm{Prob}({D_{t}}|\overline{D}_{t-1})
          \mathrm{Prob}(N|D_t),
          \end{aligned}
\end{equation}

\begin{equation}
Prob(N|C_t)=
\left\{
   \begin{array}{ll}
       \frac{\Sigma_{j=1}^{k_t}\Sigma_{l=1}^{a_t} \hat{p}_{hj}(t)\hat{p}_{jl}(1)}{\Sigma_{j=1}^{k_t}\hat{p}_{hj}(t)} &1\leq t \leq T-1\\
       \frac{\Sigma_{j=1}^{a_T} \hat{p}_{hj}(T)}{\Sigma_{l=1}^{k_T}\hat{p}_{hl}(T) } &t=T \\
   \end{array}
\right.
\end{equation}
where $a_t$ is the largest state index which satisfies
\begin{equation}
a_t=max\{k\in\{1,2,\cdots,n\}:S_{k}<(p_0-Q_0)(1+r)^t\}£¬
\end{equation}
where $S_{k}$  is the representative price level for the $kth$ state.

The proofs of two propositions can refer to the proofs of proposition 3 and 4 \cite{huang2010}.

\section{Active margin system}

\subsection{Individualized maintenance margin ratio}

It is the investo$r^{,}$s benefit to deposit the minimum margin and gain maximum leverage, this encourages the broker to permit considerable credit to increase commission income, these excessive credit transactions may have been a factor in the 1929 financial crash. This observation indicated that the more the initial margin is, the less risk the margin lending transactions is. In order to encourage the investors to deposit more initial margin, the maintenance margin ratio should be relatively smaller for those investors with more initial margin.

With the help of CPNR, it is possibility to construct individualized margin system for investors with different amounts of initial margin. For an individual investor with initial margin $Q_0+\delta P_0^{'}$, the least maintenance margin ratio $\omega^{*}$ which satisfies the constraint of CPNR can be used as the required minimal maintenance margin ratio for this inestor.

The initial margin is adequate under the constraint of CPNR if it satisfies
\begin{equation}
\frac{Q_0+\delta P_0^{'}+P_0}{P_0}\geqslant \omega,
\end{equation}
where $\omega$ is the individualized maintenance margin ratio, otherwise the investor should deposit more initial margin to fulfill the requirement of maintenance margin.

\subsection{Deduced margin system}

An artificial margin loan using cash and randomly selected stock as collateral to buy only one share of stock from 150 stocks is constructed and applied to the Chinese margin lending market in the following sections. Where only one share of stock to be purchased using an amount of cash $Q_{0}$ and a portion of the randomly selected stock $\delta$ as the collateral. Because the discount rate for stocks recommended by CSRC can not be more 70\%, so $\frac{\delta P_0^{'}}{P_{0}}\leq0.7$. The resulted margin system can be described by $(m,\delta,\omega)$, $m$ is the initial margin ratio, $\omega$ is the corresponding minimal maintenance margin ratio.

Let
\begin{eqnarray}
 m & \in &  M= \left\{ 0.01 k, k=30,31,32,\cdots, 80 \right\},   \nonumber \\
 \delta & \in &  D= \left\{ \frac{b-a}{50} k, k=0,1,2,\cdots, 50 \right\}, \nonumber \\
 w & \in &  W= \left\{ 0.01 k, k=100,101,102,\cdots, 150 \right\}, \nonumber
\end{eqnarray}
 where $a=0.01 $, $b=\frac{0.7P_0}{P_0^{'}}$.

Denoted the value of CPNR for a particular $\left(m,\delta,w\right)$ as $\text{CPNR}\left( m,\delta,w \right)$. For a fixed value of 0.05, let
\begin{equation}
\text(m_i,\delta_i,\omega_i) = \left\{ \left( m,\delta,w \right) : m\in M, \delta \in D, w \in W, 1+m \ge w, \text{CPNR}\left( m,\delta,w \right) \leq
0.05\right\},
\end{equation}
Where $i=1,2,\cdots,g$. $g$ is the total number of elements in $(m_i,\delta_i,\omega_i)$. The optimal margin system $(m^{*},\delta^{*},w^{*})$ should solve the following problem
\begin{eqnarray}
\min \sum_{i=1}^{g}(m_i-m^*)^{2}+(\delta_i-\delta^*)^{2}+(w_i-w^*)^{2}, \\  s.t. \quad (m^*,\delta^*,w^*) \in \{(m_i,\delta_i,\omega_i),i=1,2,\cdots,g\} . \nonumber
\end{eqnarray}

The resulted margin system $(m^{*},\delta^{*},w^{*})$ and the margin system recommended by CSRC are called the deduced margin system and the required margin system respectively.

\begin{table}
\caption{ Quantile analysis of the initial margin ratios under the deduced margin system.
}\label{initialtable}

\centering{
\begin{tabular*} {0.9\textwidth}{@{\extracolsep{\fill}}lllllllr}
\hline\texttt{}
		&	\multicolumn{3}{c}{} & \multicolumn{4}{c}{quantiles}	 \\	
\cline{5-8}							
	\textit{Statistics  }&	min	&	max	&	mean	&	0.70	&	0.80	&	0.90	 &	0.95	 \\ \hline
\textit{minimum} 	&   0.34  & 0.49    &0.44   & 0.45   & 0.46  &  0.47   & 0.47	 \\
\textit{maximum } 	&0.61    &0.72     &0.65    &0.66   & 0.67   & 0.69   & 0.70 	 \\
\textit{mean }	 &0.50 & 0.58  &0.58   & 0.55  &  0.56  &0.57  &0.57\\
\textit{quantiles}		\\	
0.20     &0.44    &0.55    &0.52    &0.53    &0.54    &0.54    &0.54\\
0.30     &0.47    &0.56    &0.53    &0.54    &0.54    &0.55    &0.55\\
0.40     &0.49    &0.57    &0.55    &0.55    &0.56    &0.56    &0.56\\
0.50     &0.50    &0.58    &0.56    &0.56    &0.56    &0.57    &0.58\\
0.60     &0.51    &0.60    &0.57    &0.57    &0.57    &0.58    &0.59\\
0.70     &0.52    &0.62    &0.58    &0.58    &0.59    &0.59    &0.60\\
0.80     &0.54    &0.64    &0.59    &0.59    &0.60    &0.60    &0.61\\
0.90     &0.56    &0.66    &0.60    &0.61    &0.61    &0.62    &0.62\\
0.95     &0.58    &0.67    &0.61    &0.62    &0.613    &0.64    &0.64\\

\hline
\end{tabular*}
}

\end{table}

\begin{table}
\caption{ Quantile analysis of the maintenance margin ratios under the deduced margin system.
}\label{initialtable}
\centering{
\begin{tabular*} {0.9\textwidth}{@{\extracolsep{\fill}}lllllllr}
\hline
			&	\multicolumn{3}{c}{} & \multicolumn{4}{c}{quantiles}	 \\	
\cline{5-8}							
	\textit{ Statistics  }&	min	&	max	&	mean	&	0.70	&	0.80	&	0.90	 &	0.95	 \\ \hline
\textit{minimum} 	&   1.01    &1.23    &1.03    &1.01    &1.01    &1.11   & 1.15	 \\
\textit{maximum } 	&1.10   & 1.47  &  1.27   & 1.33   & 1.36  &  1.39    &1.42	\\
\textit{mean }	 &1.03    &1.32  &1.12   & 1.18 & 1.21  &1.24 & 1.26\\
\textit{quantiles}	&		&		&		&		&		&		&	\\	
 0.20    &1.01    &1.29    &1.08    &1.14    &1.17   &1.21    &1.23\\
 0.30    &1.02    &1.30    &1.09    &1.16    &1.19    &1.23    &1.24\\
 0.40    &1.02    &1.31    &1.10    &1.18    &1.21    &1.24    &1.25\\
 0.50    &1.02    &1.32    &1.12    &1.20    &1.23    &1.26    &1.27\\
 0.60    &1.02    &1.33    &1.13    &1.21    &1.25    &1.27    &1.29\\
 0.70    &1.03    &1.34    &1.45    &1.23    &1.27    &1.28    &1.30\\
 0.80    &1.04    &1.37    &1.16    &1.25    &1.28    &1.30    &1.33\\
 0.90    &1.05    &1.39    &1.19    &1.27    &1.30    &1.32    &1.34\\
 0.95   &1.06   &1.41    &1.21    &1.29    &1.31  &1.34   &1.36\\

\hline
\end{tabular*}
}

\end{table}

\section{Empirical analysis of Chinese market}

\subsection{Design of out-of-sample test}

In this paper, the period of the loan is chosen to be 30 days, there are 800 prices just before the trading day to be used to construct the Markov chain, and there are 30 prices following this date to be used to test whether the loan is protected by the margin system, and there are 200 margin loans constructed for each stock, so the sample size need in this paper is at least 1030. If the return of the margin lending transaction after the collateral liquidated is nonnegative, it is said that the margin system passes the out-of-sample, more than that, if the frequency of negative return is not more than the specified value of CPNR (where the value is fixed at 0.05), it is said that a stock passes the test. There are 153 socks whose sample size is at least 1030 in the Shanghai 180 Stock Index, and 150 stocks have passed the out-of-sample test. In this paper the empirical investigations are limited to those 150 stocks.

\begin{table}

\caption{ Quantile analysis of the ratios of the deposited stock to the total collateral under the deduced margin system.
}\label{proportion}

\centering{
\begin{tabular*} {0.9\textwidth}{@{\extracolsep{\fill}}lllllllr}
\hline
			&	\multicolumn{3}{c}{} & \multicolumn{4}{c}{quantiles}	 \\	
\cline{5-8}							
	\textit{ Statistics  }&	min	&	max	&	mean	&	0.70	&	0.80	&	0.90	 &	0.95	 \\ \hline
\textit{minimum} 	&    0.01  &  0.40     &0.30   &  0.35   &  0.35   &  0.36    & 0.36	\\
\textit{maximum } 	&0.61  &  1.00  &  0.74  &  0.74  &  0.82  &  0.93  &  0.99	\\
\textit{mean }	 &0.40  & 0.72  &  0.51  &  0.51   & 0.51  &  0.54   & 0.58\\
\textit{quantiles}	&		&		&		&		&		&		&	\\	
 0.20   &0.25 &   0.54   & 0.45  &  0.46    &0.46  &  0.47   & 0.47\\
 0.30   & 0.34 &  0.62   & 0.47   & 0.48 &   0.48  &  0.48   & 0.50\\
 0.40   &0.37 &   0.67  &  0.49 &   0.49  &  0.50  &  0.50  &  0.52\\
 0.50   &0.39 &  0.78    &0.51    &0.51  &  0.51  &  0.52   & 0.54\\
 0.60   &0.42 &   0.84  &  0.53   & 0.52  &  0.53  &  0.54  &  0.57\\
 0.70   &0.46  &  0.92   & 0.55    &0.54   & 0.55 &   0.56  &  0.62\\
 0.80   &0.50  &  0.95  &  0.57  &  0.56 &   0.57  &  0.62   & 0.70\\
 0.90   &0.55  &  0.99   & 0.60   & 0.59 &  0.60   & 0.68  &  0.78\\
 0.95  &0.57  &  0.99   & 0.63&   0.61  &  0.63 &   0.73  &  0.82\\

\hline
\end{tabular*}
}

\end{table}

\subsection{Empirical results}

In order to investigate the distributional properties of those variables from the view point of the whole market, a quantile analysis is applied to the observed variables of the 150 stocks. There are 12 statistics observed from those variables of each stock, including the minimum value, the maximum value, mean, and 20\%,30\%,40\%,50\%,60\%,70\%,80\%,90\%,95\% quantiles respectively.

There are 150 observations for every statistic. From the Table 1 and 2, it can be found that the mean of those observed initial margin ratios is 58\%, and the average value of those maintenance margin ratios is 112\%, which is smaller than the required minimal maintenance margin ratio 130\%.

The proportion of stock in the initial margin is
\begin{equation}
p=\frac{\delta P_0^{'}}{\delta P_0^{'} + Q_0},
\end{equation}
which is reported in Table 3, the average value of those observed proportions is 51\%. The means of those observed 9 quantiles are distributed from 45\% to 63\%, which indicates that the proportion of stock in the initial margin is very big.

Costs of margin loans under the deduced and the required margin systems are reported in Table 4. In the first row, the values for each statistic are the observations under the deduced margin system, and the second row are those observations given by the required margin system, the different between those two 95\% quantiles indicate that the average cost of each margin loan under the deduced margin system is about 4\% less than the average cost under the required margin system.

 \begin{table}[h]
\caption{Costs of margin loans under the deduced and the required margin systems.}\label{loancost}
\centering{
\begin{tabular*}{\textwidth}{@{\extracolsep{\fill}}llllllllr}
\hline
	&	\multicolumn{3}{c}{}	& \multicolumn{4}{c}{quantiles}	& \\
\cline{5-8}	
	\textit{ Statistics  }&	min	&	max	&	mean	&	0.70 	&	0.80 	&	0.90 	 &	0.95 	  & RD 	 \\
\hline
\textit{minimum}	&	1.45   &62.91    &5.93    &6.28    &7.99    &10.39   &12.90   &-0.04			 \\
                 	&	1.32   &51.25    &5.87   & 6.24 &   8.20   & 10.05 & 13.45	 \\
\textit{maximum}	&	1.98   &269.09  & 26.00  & 28.60 &  36.92  & 48.08  & 64.98  &-0.31			 \\
	               &	2.31 & 124.68   &18.07  & 19.10 &  24.99 &  34.63  &45.03	 \\
\textit{mean}	   &	1.73  & 83.75  &  9.56 &   11.02   &13.81   &18.46  & 23.76 & -0.14	 \\
	               &	1.85   &88.25 & 9.15  & 9,99  & 12.95  & 16.57 &  20.37 	 \\
\textit{quantiles}	&		&		&		&		&		&		&		&		 \\
0.20    &1.53   &67.99  & 6.87   & 7.54   & 9.22 &11.50  & 16.34  & 0.07\\
       &1.62   &74.73    &7.49   &8.33   &10.15   &12.92  &17.51\\
0.30   & 1.60   &73.38   &7.21  & 7.83   &10.01  & 12.02  & 16.84   &0.09\\
      & 1.68   &78.17    &7.85    &8.80    &10.97   &13.22   &18.29\\
0.40   & 1.64   &79.67  & 7.60   & 8.03   &10.84 & 12.80  & 17.25   &0.09\\
       &1.74  &83.90    &8.25    &9.17   &11.60   &13.94  &18.77\\
0.50   & 1.66  & 82.93   & 7.97  & 8.28  & 11.59   &13.49   &19.71  &-0.03\\
       &1.78   &88.31    &8.67    &9.64   &12.50   &14.98   &19.12\\
0.60    &1.69  & 85.35  & 8.40   &8.76  & 12.50  & 15.14  & 20.92  & -0.01\\
       &1.82   &92.79    &9.13   &10.13  &13.10   &15.89   &20.82\\
0.70    &1.73   &88.69   & 9.00   &9.29   &13.12  & 16.00  & 26.40  & -0.15\\
       &1.88   &95.32    &9.63   &10.41   &14.24   &17.31   &22.56\\
0.80    &1.76  & 90.99   &10.71   &11.91  & 15.64   &22.58   &31.76  &-0.21\\
       &1.93  &101.27  &10.33   &10.92   &15.07   &19.24   &25.00\\
0.90    &1.80  & 95.48  & 15.98  & 18.93  & 25.51  & 36.03 & 41.88  &-0.21\\
       &2.01   &104.69   &11.99   &12.42   &16.42   &24.15   &32.91\\
0.95   &1.85   &108.21  &19.75  & 23.48   &14.57   &39.95   &58.24  &-0.4\\
      &2.08  &109.22  &13.43  & 14.58   &17.36   &25.80   &34.79\\

\hline
\end{tabular*}
}
\end{table}

\begin{table}
\caption{Numbers of margin calls under the required and the deduced margin systems.}\label{callnumber}
\centering{
\begin{tabular*}{\textwidth}{@{\extracolsep{\fill}}lllllllllr}
\hline
	&	\multicolumn{3}{c}{}	& \multicolumn{6}{c}{quantiles}	\\

\cline{5-10}	
	&	min	&	max	&	mean	&	0.30 	&	0.50 	&	0.80 	&	0.90 	 &	 0.95 	&	0.99 	\\ \hline
\textit{Required}	&	0.00 &75.00  &19.49  &7.00 &18.00   &35.00 &  41.00 &51.00 &  54.00 	 \\
\textit{Deduced}	&	0  &  50.00  & 5.29  &0.00 & 0.00   &9.00 &  21.00 & 27.00 & 39.00 	 \\

\hline
\end{tabular*}
}
\end{table}

 From Table 5, we can find that the mean of the 150 observed numbers of margin calls is 19.49 under the required margin system, and the mean of the 150 observed numbers of margin calls is 5.29 under the deduced margin system, the number of margin calls under the deduced margin system is much less than the number under the required margin system.

\section{Conclusions and discussions}

This paper constructs an active margin system for the margin loans using cash and randomly selected stock as collateral, the active system focuses on three aspects: the initial margin requirement, maintenance margin requirement and mandatory liquidation, which is the line defence against the risk of the margin lending. The conditional probability of negative return (CPNR) that the broker gets by mandatory liquidation is used as a risk metric. With the help of CPNR, it is possibility to construct individualized margin system for investors with different amounts of initial margin. The deduced margin system is dynamic, which responses to the changes of stock market.

The performance of individualized maintenance margin is investigated through the deduced margin system, which is chosen from the indifference set of margin system by a least squares method.

The empirical results show that the average ratio of the initial margin is 58\% under the deduced margin system, which is larger than 50\% recommended by CSRC, the average ratio of maintenance margin is 112\% under the deduced margin system, which is less than 130\% recommended by CSRC, the average proportion of stock in the initial margin is around 51\%, the number of margin calls for each stock within 200 margin loans is around 19.49 and 5.29 under the required margin system and the deduced margin system respectively, the cost of each margin loan is averagely 9.56 yuan and 9.15 yuan under the required margin system and the deduced margin system respectively. In other words, the frequency of margin call is reduced by 73\% with less 4\% cost under the deduced margin system. Those observations indicate that CPNR can be used to measure the risk faced by brokers, and the deduced margin system is an operational active margin system. This paper does not give the performance of margin system constructed in this paper, it will be left for further research.

\section*{Acknowledgments}
 {Guanghui Huang is supported by the Fundamental Research Funds for the Central Universities of China under Grant CDJZR10 100 007.}

\bibliographystyle{elsarticle-harv}



\end{document}